\documentclass[11pt]{article}
\usepackage{times}
\usepackage{amsmath}
\usepackage{amsfonts}
\usepackage{latexsym}
\usepackage{tikz,algonjk}
\usepackage{graphicx}
\usepackage{url,hyperref}

\advance \topmargin by -1.0in
\advance \oddsidemargin by -0.8in
\newcommand{\myparskip}{3pt}
\parskip \myparskip

\newtheorem{lemma}{Lemma}[section]
\newtheorem{theorem}[lemma]{Theorem}

\newtheorem{prop}[lemma]{Proposition}

\newtheorem{conjecture}{Conjecture}

\newenvironment{proof}{\vspace{-0.15in}\noindent{\bf Proof:}}%
        {\hspace*{\fill}$\Box$\par}
        {\hspace*{\fill}$\Box$\par\vspace{4mm}}
\newenvironment{proofof}[1]{\smallskip\noindent{\bf Proof of #1.}}%
        {\hspace*{\fill}$\Box$\par}

\newcommand{\etal}{{\em et al.}\ }

\newcommand{\eps}{\varepsilon}

\newcommand{\ceil}[1]{\lceil #1 \rceil}
\newcommand{\floor}[1]{\lfloor #1 \rfloor}
\newcommand{\opt}{\text{\sc OPT}}

\newcommand{\Exp}[1]{\mathbb{E}[#1]}

\newcommand{\Rev}{Revenue}
\DeclareMathAlphabet{\mathpzc}{OT1}{pzc}{m}{it}
\newcommand{\script}[1]{\mathcal{#1}}

\begin{document}

\title{Algorithms for Secretary Problems on Graphs and Hypergraphs}
\author{
Nitish Korula\thanks{Dept.\ of Computer Science, University of Illinois, Urbana, IL 61801. This work was done while the author was at Google Inc., NY. {\tt nkorula2@uiuc.edu}}
\and
Martin P\'al\thanks{Google Inc., 76 9th Avenue, New York, NY 10011. {\tt mpal@google.com}}
}

\maketitle

\begin{abstract}
  We examine several online matching problems, with applications to
  Internet advertising reservation systems. Consider an edge-weighted
  bipartite graph $G$, with partite sets $L,R$. We develop an
  $8$-competitive algorithm for the following secretary problem:
  Initially given $R$, and the size of $L$, the algorithm receives the
  vertices of $L$ sequentially, in a random order.  When a vertex $l
  \in L$ is seen, all edges incident to $l$ are revealed, together
  with their weights. The algorithm must immediately either match $l$
  to an available vertex of $R$, or decide that $l$ will remain
  unmatched.

  In \cite{DimitrovPlaxton}, the authors show a 16-competitive
  algorithm for the transversal matroid secretary problem, which is
  the special case with weights on vertices, not edges.
  (Equivalently, one may assume that for each $l \in L$, the weights
  on all edges incident to $l$ are identical.) We use a similar
  algorithm, but simplify and improve the analysis to obtain a better
  competitive ratio for the more general problem. Perhaps of more
  interest is the fact that our analysis is easily extended to obtain
  competitive algorithms for similar problems, such as to find
  disjoint sets of edges in hypergraphs where edges arrive online. We
  also introduce secretary problems with adversarially chosen
  \emph{groups}.

  Finally, we give a $2e$-competitive algorithm for the secretary
  problem on graphic matroids, where, with edges appearing online, the
  goal is to find a maximum-weight acyclic subgraph of a given graph.
\end{abstract}

\setcounter{page}{0}
\thispagestyle{empty}
\clearpage

\section{Introduction}
\label{sec:intro}

Many optimization problems of interest can be phrased as picking a
maximum-weight independent subset from a ground set of elements, for a
suitable definition of independence. A well-known example is the
(Maximum-weight) Independent Set problem on graphs, where we wish to
find a set of vertices, no two of which are adjacent. A more tractable
problem in this setting is the Maximum-weight Matching problem, in
which we wish to find a set of edges such that no two edges share an
endpoint. This notion of independence can be naturally extended to
hypergraphs, where a set of hyperedges is considered independent if no
two hyperedges share a vertex.

In the previous examples, independent sets are characterized by
forbidding certain \emph{pairs} of elements from the ground set.  A
somewhat related, but different notion of independence comes from the
independent sets of a matroid. For example, in the uniform matroid of
rank $k$, any set of at most $k$ elements is independent. For graphic
matroids, a set of edges in an undirected graph is independent if and
only if it does not contain a cycle; the optimization goal is to find
a maximum-weight acyclic subgraph of a graph $G$. In transversal
matroids, a set of left-vertices of a bipartite graph is independent
if and only if there is a matching that matches each vertex in this
set to some right-vertex.

In many applications, the elements of the ground set and their weights
are not known in advance, but arrive online one at a time. When an
item arrives, we must immediately decide to either irrevocably accept
it into the final solution, or reject it and never be able to go back
to it again. We will be interested in competitive analysis, that is,
comparing the performance of an online algorithm to an optimal offline
algorithm which is given the whole input in advance.  In this setting,
even simple problems like selecting a maximum-weight element become
difficult, because we do not know if elements that come in the future
will have weight significantly higher or lower than the element
currently under consideration. If we make no assumptions about the
input, any algorithm can be fooled into performing arbitrarily poorly
by ofering it a medium-weight item, followed by a high-weight item if
it accepts, and a low-weight item if it rejects.  To solve such
problems, which frequently arise in practice, various assumptions are
made. For instance, one might assume that weights are all drawn from a
known distribution, or (if independent sets may contain several
elements) that the weight of any single element is small compared to
the weight of the best independent set.

One useful assumption that can be made is that the elements of the
ground set appear in a random order. The basic problem in which the
goal is to select the maximum-weight element is well known as the
\emph{Secretary Problem}. It was first published by Martin Gardner in
\cite{Gardner}, though it appears to have arisen as folklore a decade
previously \cite{History}. An optimal solution is to observe the first
$n/e$ elements, and select the first element from the rest with weight
greater than the heaviest element seen in the first set; this
algorithm gives a $1/e$ probability of finding the heaviest element,
and has been attributed to several authors (see \cite{History}).

Motivated by this simple observation, several results have appeared
for more complex problems in this random permutation model; these are
often called secretary-type problems. Typically, given a random
permutation of elements appearing in an online fashion, the goal is to
find a maximum-weight independent set. For example, Kleinberg
\cite{KleinbergKSec} gives a $1+O(1/\sqrt{k})$-competitive algorithm
for the problem of selecting at most $k$ elements from the set to
maximize their sum. Babaioff \etal \cite{KnapsackSec} give a
constant-competitive algorithm for the more general Knapsack secretary
problem, in which each element has a size and weight, and the goal is
to find a maximum-weight set of elements whose total size is at most a
given integer $B$.

Babaioff \etal \cite{MatroidSec} had earlier introduced the so-called
\emph{matroid secretary problem}, and gave an $O(\log k)$-competitive
algorithm to find the max-weight independent set of elements, where
$k$ is the rank of the underlying matroid. A $16$-competitive
algorithm was also given in \cite{MatroidSec} for the special case of
graphic matroids; this was based on their $4d$-competitive algorithm
algorithm for the important case of \emph{transversal matroids}, where
$d$ is the maximum degree of any left-vertex. Recently, Dimitrov and
Plaxton \cite{DimitrovPlaxton} improved the latter to a ratio of $16$
for all transversal matroids. A significant open question is whether
there exists a $O(1)$-competitive algorithm for general matroids, or
for other secretary problems with non-matroid constraints.

\bigskip 
These secretary-type problems arise in many practical situations where
decisions must be made in real-time without knowledge of the future,
or with very limited knowledge. For example, a factory needs to decide
which orders to fulfil, without knowing whether more valuable orders
will be placed later. Buyers and sellers of houses must decide whether
to go through with a transaction, though they may receive a better
offer in a week or a month. Below, we give an example from online
advertising systems, which we use as a recurring motivation through
the paper.

Internet-based systems are now being used to sell advertising space in
other media, such as newspapers, radio and television broadcasts,
etc. Advertisers in these media typically plan advertising campaigns
and reserve slots well in advance to coincide with product launches,
peak shopping seasons, or other events. In such situations, it is
unreasonable to run an auction immediately before the event to
determine which ads are shown, as is done for sponsored search and
other online advertising.

Consider an automatic advertising reservation system, in which the
seller controls a number of \emph{slots}, each representing a position
in which an advertisement (hereafter \emph{ad}) can be published.
Advertisers/Bidders appear periodically, and report which slots they
would like to place an ad in, and how much they are willing to pay for
each slot. When an advertiser reports a bid, the system must
immediately decide whether or not to accept it; if a bid is accepted,
the ad \emph{must} be placed in the corresponding slot, and if not,
the ad is permanently rejected.  Note that in disallowing the removal
of an accepted ad, our model differs significantly from that of
\cite{OnlineBump}, in which the seller can subsequently remove an
accepted ad if he makes a compensatory payment to the advertiser.

We model this system as as an online edge-weighted matching problem on
a bipartite graph $G(L \cup R, E)$: the vertices of set $R$ correspond
to the set of slots, and those of set $L$ to the ads. For each vertex
$l \in L$, its neighbors in $R$ correspond to the slots in which ad
$l$ can appear, and the weight of edge $(l,r)$ is the amount the
advertiser is willing to pay if $l$ appears in slot $r$. Initially,
the seller knows the set of slots $R$; vertices of $L$ appear
sequentially in a random order, as advertisers bid on slots. When a
vertex $l \in L$ is seen, all the edges from $l$ to $R$ are revealed,
together with their weights; the seller must immediately decide
whether to accept ad $l$, and if so, which of the relevant slots to
place it in. The seller's goal, obviously, is to maximize his
revenue. Subsequently, we refer to this problem as Bipartite
Vertex-at-a-time Matching (BVM). We describe our results for BVM and
other problems below.

\subsection{Results and Outline}
Recall that the elements of a transversal matroid are one partite set
$L$ (subsequently referred to as the \emph{left vertices}) of a
bipartite graph, and a set of vertices $S \subseteq L$ is independent
if the graph constains a perfect matching from $S$ to the other
partite set. That is, the transversal matroid secretary problem is
equivalent to the special case of BVM in which all edges incident to
each $l \in L$ have the same weight.  (Equivalently, the weights are
on vertices of $L$ instead of edges.) In Section~\ref{sec:BVM}, we
give a simpler and tighter analysis for an algorithm essentially
similar to that of Dimitrov and Plaxton \cite{DimitrovPlaxton} for
transversal matroids; this allows us to improve the competitive ratio
from 16 to 8, even for the more general BVM problem.

In addition to an improved ratio, our methods are of interest as they
appear robust to changes in the model and can be naturally applied to
more general problems. We illustrate this in
Section~\ref{sec:hypergraphs} by extending our algorithms to
hypergraph problems, with applications to more complex advertising
systems in which advertisers desire \emph{bundles} of slots, as
opposed to a single slot. In particular, we obtain
constant-competitive algorithms for finding independent edge sets in
hypergraphs of constant edge-size.

We also introduce secretary problems with \emph{groups}, to model
applications in which we do not see a truly random permutation of
elements. We assume that an adversary can group the elements
arbitrarily, but once the groups are constructed, they appear in
random order. When a group appears, the algorithm can see all the
elements in the group. We discuss this idea further in
Section~\ref{sec:groups}.

Finally, in Section~\ref{sec:graphic}, we obtain a simple
$2e$-competitive algorithm for the problem of finding independent
edge-sets in graphic matroids, improving the ratio of 16 from
\cite{MatroidSec}.

The majority of our algorithms follow the ``sample-and-price'' method
common to many solutions to secretary problems. That is, we look at a
random sample of elements containing a constant fraction of the input,
and use the values observed to determine \emph{prices} or thresholds.
In the second half, we accept an element if its weight/value is above
the given price. For instance, in the optimal solution to the original
secretary problem, the price is set to be the highest value seen in
the first $1/e$ fraction of the input, and we accept any element from
the remaining set with value greater than this price.

\section{The Bipartite Vertex-at-a-time Matching Problem}
\label{sec:BVM}

Recall that in the BVM problem, the algorithm is initially given one
partite set $R$ of a bipartite graph $G(L \cup R, E)$, together with
the size of the other partite set $L$. The algorithm sees the vertices
of $L$ sequentially, in a random order.  When a vertex $l \in L$ is
seen, all edges incident to $l$ are revealed, together with their
weights. The algorithm must immediately either match $l$ to an
available vertex of $R$, or decide that $l$ will remain permanently
unmatched. In this section, we show that an algorithm based on that of
\cite{DimitrovPlaxton} gives a competitive ratio of 8 for this problem.
Before presenting the algorithm for BVM, we describe a closely related
algorithm {\sc Simulate} that is easier to analyze, and then show that
our final algorithm does at least as well as {\sc Simulate}.

\noindent Let {\sc Greedy} denote the following greedy algorithm for
the offline Edge-weighted bipartite matching problem:

\begin{algo}
\underline{\sc Greedy($G(L \cup R, E)$)}:\\
  Sort edges of $E$ in decreasing order of weight. \\
  Matching $M \leftarrow \emptyset$ \\
  For each edge $e \in E$, in sorted order \+ \\
    If $M \cup e$ is a matching: \+ \\
      $M \leftarrow M \cup e$  \- \- \\
  Return $M$.
\end{algo}

Let $w(F)$ denote the weight of a set of edges $F$, and $\opt$ denote
the weight of an optimum (max-weight) matching on $G$. It is easy to
see the following proposition, that {\sc Greedy} is a 2-approximation.
\begin{prop}
  $w(M) \ge \opt/2$.
\end{prop}

We now describe the algorithm {\sc Simulate}, which we use purely to
analyze our final algorithm for BVM.

\begin{algo}
\underline{\sc Simulate}:\\
  Sort edges of $G(L \cup R,E)$ in decreasing order of weight. \\
  $M_1, M_2 \leftarrow \emptyset$ \\
  Mark each vertex $l \in L$ as unassigned.\\
  For each edge $e = (l,r) \in E$, in sorted order \+ \\
    If $l$ is unassigned {\bf AND} $M_1 \cup e$ is a matching: \+ \\
      Mark $l$ as assigned \\
      Flip a coin with probability $p$ of heads \\
      If heads, $M_1 \leftarrow M_1 \cup e$ \\
      Else $M_2 \leftarrow M_2 \cup e$  \- \- \\
  $M_3 \leftarrow M_2$ \\
  For each vertex $r \in R$ \+ \\
    If $r$ has degree $> 1$ in $M_3$ \+ \\
      Delete all edges incident to $r$ from $M_3$. \- \-
\end{algo}

Say that an edge $e$ is \emph{considered} by {\sc Simulate} if we flip
a coin and assign $e$ to either $M_1$ or $M_2$. We make two
observations about {\sc Simulate}: Once any edge incident to a vertex
$l \in L$ has been considered, no other edge incident to $l$ will be
considered later. Second, once an edge incident to $r \in R$ has been
added to $M_1$, no subsequent edge incident to $r$ will be
considered. (Note that multiple edges incident to $r$ might be
considered until one of these edges is added to $M_1$.)

Observe that from our description of {\sc Simulate}, $M_1$ is a
matching, but $M_2$ may not be, as a vertex $r \in R$ may be incident
to multiple edges of $M_2$. Hence, we have a final pruning step in
case there are multiple edges incident to the same vertex of $R$; this
gives us a matching $M_3$. We now prove three statements about {\sc
  Simulate}, and later show that the matching returned by our online
algorithm is at least as good as $M_3$.

\begin{prop}\label{prop:wM1}
  $\Exp{w(M_1)} \ge p \opt/2$.
\end{prop}
\begin{proof}
  {\sc Simulate} tosses a coin (at most) once for each vertex in $L$;
  $M_1$ is precisely the matching one would obtain from running {\sc
    Greedy} on $L'\cup R$, where $L'$ denotes the vertices which came
  up heads. (If the coin for a vertex comes up tails, this vertex has
  no effect on $M_1$.) If $\opt'$ denotes the weight of an optimum
  matching on $L' \cup R$, it is easy to see that $\Exp{\opt'} \ge p
  \opt$, and hence that $\Exp{w(M_1)} \ge p \opt/2$.
\end{proof}


\begin{lemma}\label{lem:wM2}
  $\Exp{w(M_2)} \ge (1-p) \opt/2$.
\end{lemma}
\begin{proof}
  Consider any history of coin tosses in which an arbitrary edge $e$
  is being considered, and we are about to flip a coin to determine
  whether $e$ is added to $M_1$ or $M_2$. Its expected contribution to
  $M_1$ is $p w(e)$, and to $M_2$, is $(1-p) w(e)$. This holds for
  each edge $e$ and any history in which $e$ can contribute to the
  weight of $M_1$ or $M_2$; hence $\Exp{w(M_2)} = \frac{(1-p)}{p}
  \Exp{w(M_1)}$, completing the proof.
\end{proof}

\begin{lemma}\label{lem:wM3}
  $\Exp{w(M_3)} \ge \frac{p^2(1-p)}{2} \opt$.
\end{lemma}
\begin{proof}
  For each vertex $v \in R$, let $\Rev_2(v)$ denote the revenue earned
  by vertex $v$ in $M_2$, which we define as the sum of the weights of
  edges in $M_2$ incident to $v$. (Hence, $\sum_v \Rev_2(v) =
  w(M_2)$.)  For each edge $e$ incident to $v$, let
  $\Exp{\Rev_2(v)|e}$ denote the expected revenue earned by $v$ in
  $M_2$, conditioned on the fact that $e$ is the first edge incident
  to $v$ selected by {\sc Simulate} for $M_2$.  It is easy to see that
  $\Exp{\Rev_2(v)|e} \le w(e)/p$, by considering how $v$ can earn
  revenue: If the next edge incident to $v$ considered by {\sc
    Simulate} is added to $M_1$ (which happens with probability $p$),
  then $v$ earns precisely $w(e)$, as no later edge incident to $v$
  can ever be considered. In general, if $v$ is incident to $i$ edges
  in $M_2$, the revenue it earns is at most $i w(e)$, and the
  probability of this event is at most $(1-p)^{i-1}\cdot p$; this is
  because the next $i-1$ edges incident to $v$ that are considered
  must be added to $M_2$, and the $i$th edge is added to
  $M_1$. Therefore, $\Exp{\Rev_2(v) | e} \le w(e) \sum_{i=1}^\infty i
  \cdot p (1-p)^{i-1} = w(e)/p$.

  Similarly, for each vertex $v \in R$, let $\Rev_3(v)$ denote the
  revenue earned by vertex $v$ in $M_3$, which is the weight of the
  (at most one) edge incident to $v$ in $M_3$. Let $\Exp{\Rev_3(v)|e}$
  denote the expected revenue earned by $v$, conditioned on $e$ being
  the first edge incident to $v$ added to $M_2$. With probability $p$,
  the next considered edge incident to $v$ is added to $M_1$, and
  hence $v$ has degree 1 in $M_2$. Therefore, $\Exp{\Rev_3(v)|e} \ge p
  w(e)$, and so $\Exp{\Rev_3(v)|e} \ge p^2 \Exp{\Rev_2(v)|e}$; it
  follows that $\Exp{w(M_3)} \ge p^2 \Exp{w(M_2)} = \frac{p^2
    (1-p)}{2} \opt$.
\end{proof}

Before describing our final algorithm for EBP, we show that the
matching returned by an intermediate algorithm {\sc SampleAndPermute} is
at least as good as $M_3$, which implies that we have a
$\frac{2}{(1-p)p^2}$-competitive algorithm: setting $p = 2/3$, we get
a $13.5$-competitive algorithm. However, our pruning step allows us to
take an edge for $M_3$ only if its right endpoint has degree 1; a more
careful pruning step allows more edges in the matching. We use this
fact to give a tighter analysis for the next algorithm, obtaining a
competitive ratio of 8.

\begin{figure}
\begin{algo}
  \underline{\sc SampleAndPermute($G(L \cup R, E)$)}:\\
  $L' \leftarrow \emptyset$ \\
  For each $l \in L$: \+ \\
    With probability $p$, $L' \leftarrow L' \cup \{l\}$ \- \\
  $M_1 \leftarrow$ {\sc Greedy($G[L' \cup R]$)}. \\
  For each $r \in R$: \+ \\
    Set $price(r)$ to be the weight of the edge incident to $r$ in
    $M_1$. \- \\
  $M, M_2 \leftarrow \emptyset$ \\
  For each $l \in L - L'$, in random order: \+ \\
    Let $e = (l,r)$ be the highest-weight edge such that $w(e) \ge
    price(r)$ \\
    Add $e$ to $M_2$.\\
    If $M \cup e$ is a matching, add $e$ to $M$. \- 
\end{algo}
\end{figure}

Note that the matching $M_1$ in {\sc SampleAndPermute} is precisely
the same as $M_1$ from {\sc Simulate}; intuitively, in the former, we
toss all the coins at once and run {\sc Greedy}, while in the latter,
we toss coins while constructing the Greedy Matching. (More precisely,
the two algorithms to generate the matchings are equivalent.)
Similarly, the ``matching'' $M_2$ in this algorithm is essentially
$M_2$ from {\sc Simulate}. The difference between the two algorithms
is in the pruning step: To construct $M_3$ in {\sc Simulate}, we
delete all edges incident to any vertex $r \in R$ with degree greater
than 1; in {\sc SampleAndPermute}, we add to $M$ the first such edge
seen in our permutation of $L - L'$. It follows immediately from
Lemma~\ref{lem:wM3} that $\Exp{w(M)} \ge p^2(1-p) \opt/2$, but
accounting for the difference in pruning allows the following tighter
statement, which we prove in the appendix.

\begin{lemma}\label{lem:tighterBound}
  $\Exp{w(M)} \ge \frac{p (1-p)}{2} \opt$.
\end{lemma}

We now present our final algorithm, a trivial modification of {\sc
  SampleAndPermute} for the online BVM problem.

\vspace{-0.25in}
\begin{figure}[h]
\begin{algo}
  \underline{\sc SampleAndPrice($|L|, R$)}\\
  $k \leftarrow Binom(|L|, p)$ \\
  Let $L'$ be the first $k$ vertices of $L$.\\
  $M_1 \leftarrow$ {\sc Greedy($G[L' \cup R]$)}. \\
  For each $r \in R$: \+ \\
    Set $price(r)$ to be the weight of the edge incident to $r$ in
    $M_1$. \- \\
  $M \leftarrow \emptyset$ \\
  For each subsequent $l \in L - L'$, : \+ \\
    Let $e = (l,r)$ be the highest-weight edge such that $w(e) \ge
    price(r)$ \\
    If $M \cup e$ is a matching, accept $e$ for $M$. \-
\end{algo}
\end{figure}
\vspace{-0.1in}

As the input to {\sc SampleAndPrice} is a random permutation, $L'$ is
a subset of $L$ in which each vertex of $L$ is selected with
probability $p$; it is easy to see that this algorithm is equivalent
to {\sc SampleAndPermute}. Therefore, $\Exp{w(M)} \ge \frac{p
  (1-p)}{2} \opt$; setting $p = 1/2$ implies that the expected
competitive ratio is 8.

\section{Independent Edge Sets in Hypergraphs}
\label{sec:hypergraphs}

In the Hypergraph Edge-at-a-time Matching (HEM) problem, we are
initally given the vertex set of a hypergraph; subsequently,
hyperedges appear in a random order. When an edge (together with its
weight) is revealed, the algorithm must immediately decide whether or
not to accept it; as before, the goal is for the algorithm to select a
maximum-weight set of disjoint edges. For arbitrary hypergraphs, one
can observe that even the offline version of this problem is
NP-Complete (and also hard to approximate) via an easy reduction from
the Independent Set problem.  However, the difficulty is related to
the size of the hyperedges; if all edges contain only 2 vertices, for
instance, then we are simply trying to find a matching in a (possibly
non-bipartite) graph. (Even in this special case, the problem is of
interest in an online setting.) Let $d$ denote the maximum size of an
edge in the hypergraph. 

We provide an $O(d^2)$-competitive algorithm for the HEM problem by
solving the more general Hypergraph Vertex-at-a-time Matching (HVM)
problem, described as follows: We are initally given a subset $R$ of
the vertex set of a hypergraph. The remaining vertices $L$ arrive
online; each edge of the hypergraph is constrained to contain exactly
one vertex of $L$, together with some vertices of $R$. The vertices of
$L$ appear online in a random order; when $l \in L$ is revealed, the
algorithm also sees all edges incident to $l$, together with their
weight. At this point, the algorithm must immediately decide whether
or not to accept some edge containing $l$, and if so, which edge;
again, the goal is for the algorithm to select a maximum-weight set of
disjoint edges. Here, let $d$ denote the maximum number of vertices of
$R$ contained in a single edge (so the largest edge has $d+1$
vertices). First, we observe that the HEM problem with edge size $d$
reduces to the HVM problem with edge size $d+1$: Let $R$ be the
vertex set of the original hypergraph, and add one vertex to $L$ for
each original edge. An edge of the new hypergraph consists of an old
edge, together with the corresponding vertex of $L$. Clearly,
observing a random permutation of $L$ together with the incident edges
is equivalent to a random permutation of the edge set of the original
hypergraph.  Also, notice that the the BVM problem of
Section~\ref{sec:BVM} is simply the special case of HVM when
$d=1$. (See Figure 1 at the end of this section.)

These hypergraph problems capture the notion of demand
\emph{bundles}. For instance, in ad reservation systems, advertisers
rarely make reservations for a single ad at a time; they are more
likely to plan advertising campaigns involving multiple individual
ads. In many campaigns, advertisers create various ads which are
related to and complement or reinforce each other; these advertisers
might be interested in acquiring a bundle or set of slots for this
campaign. They submit to the reservation system the bundles they are
interested in, together with the price they are willing to pay; the
system must either accept a request for an entire bundle or reject it,
as it does not receive revenue for providing the advertiser with a
part of the bundle. If each advertiser submits a request for a single
bundle, we obtain the HEM problem with vertex set corresponding to the
set of slots. More generally, an advertiser may submit a request for
\emph{one of} a set of bundles, together with a price for each
bundle. (For example, an advertiser might want an ad to appear in
any three out of four local newspapers.) This leads to the HVM
problem, with vertex set $L$ corresponding to the set of advertisers,
and set $R$ to the set of slots: We receive a random permutation of
advertisers, and each advertiser informs us of the bundles she is
interested in, together with a price for each bundle.

Let {\sc Greedy} denote the offline algorithm for HVM that sorts edges
in decreasing order of weight, and selects an edge if it is disjoint
from all previously selected edges. For ease of exposition, we
subsequently assume that the hypergraph is $(d+1)$-uniform; that is,
that each edge contains exactly $d$ vertices of $R$ together with one
vertex of $L$.

\begin{prop}
  {\sc Greedy} returns a $(d+1)$-approximation to the maximum-weight
  disjoint edge set.
\end{prop}

We again define an algorithm {\sc Simulate}, as in
Section~\ref{sec:BVM}:

\begin{algo}
  Sort edges of $E$ in decreasing order of weight.\\
  Mark each vertex $l \in L$ as unassigned.\\
  $M_1, M_2 \leftarrow \emptyset$\\
  For each edge $e \in E$ in sorted order: \+ \\
    Let $l$ be the vertex of $L$ in $e$ \\
    If $l$ is unassigned {\bf AND} $e$ is disjoint from $M_1$: \+ \\
      Mark $l$ as assigned. \\
      Flip a coin with probability $p$ of heads\\
      If heads, add $e$ to $M_1$ \\
      If tails, add $e$ to $M_2$ \- \- \\      
  $M_3 \leftarrow \emptyset$ \\
  For each $e \in M_2$: \+ \\
    Add $e$ to $M_3$ if $e$ is disjoint from the rest of $M_2$. 
\end{algo}

As before, we let $w(F)$ denote the weight of an edge set $F$. The
proofs of the following two propositions are exactly analogous to
Proposition~\ref{prop:wM1} and Lemma~\ref{lem:wM2}.
\begin{prop}
  $\Exp{w(M_1)} \ge p \cdot \opt/(d+1)$.
\end{prop}
\begin{prop}
  $\Exp{w(M_2)} \ge (1-p) \opt/(d+1)$.
\end{prop}

It is now slightly more complex to bound the weight of $M_3$ than it
was for the BVM problem; for BVM, the set of edges in $M_2$ incident
to $v \in R$ interfere only with each other, but in the hypergraph
version, edges $e_1$ and $e_2$ might not intersect, though they may
both intersect $e_3$, and hence all of $e_1, e_2, e_3$ will have to be
deleted. However, we can use a similar intuition: In BVM, we charge
all edges of $M_2$ incident to $v$ to the heaviest such edge; in
expectation, each edge is charged a constant number of times. For the
HVM problem, we charge all the edges in a ``connected component'' to
the heaviest edge in the component, and argue that (with a suitable
choice of $p$) the average size of the components is small. More
formally, we prove the following lemma: 

\begin{lemma}\label{lem:HVM}
  Setting $p = 1 - 1/2d$, $\Exp{w(M_3)} \ge \frac{\opt}{12 d(d+1)}$.
\end{lemma}
\smallskip
\begin{proof}
  Construct an auxiliary directed graph $F$ as follows: For each $e
  \in M_2$, add a corresponding vertex $v_e$ to $F$. If $e'$ is the
  heaviest edge in $M_2$ that intersects $e$, add a directed arc from
  $v_e$ to $v_{e'}$ to $F$. (If $e$ itself is this heaviest edge,
  $v_e$ has no out-neighbors.) Note that the graph $F$ is obviously a
  forest. For each $e \in M_2$, if $v_e$ is not the root of its tree
  in $F$, we define $\Rev_2(e)$ to be 0, and if it is the root, we set
  $\Rev_2(e)$ to be the weight of all edges of $M_2$ in the
  tree. Clearly, $\sum_e \Rev_2(e) = w(M_2)$.

  We define $\Rev_3(e)$ to be equal to the weight of $e$ iff $e$ is an
  edge in $M_2$ that does not intersect any other such edge. (In which
  case, it follows that $v_e$ is the root of its tree.) We prove that
  $\Exp{\Rev_3(e)} \ge \frac{\Exp{\Rev_2(e)}}{6}$, which proves the
  lemma, since $\sum_e \Rev_3(e) = w(M_3)$.

  First, note that the probability that any edge $e$ added to $M_2$
  intersects an edge added later is at most $1/2$: For each vertex $u$
  of $R$ contained in $e$, the probability that $e$ intersects a later
  edge because of $u$ is at most $1/2d$, as with probability $1 -
  1/2d$, the next edge containing $u$ considered by {\sc Simulate}
  will be added to $M_1$. As $e$ contains only $d$ vertices in $R$,
  the desired probability is at most $1/2$. (Every vertex of $L$ is
  incident to at most one edge in $M_2$, and so $e$ cannot intersect
  any other edge through its vertex in $L$.) It follows that the
  probability that any $v_e \in F$ has a child is at most $1/2$. We
  also count the expected number of children of $v_e$; the edge
  corresponding to each child of $v_e$ must share some vertex with
  $e$, and the expected number of children through a particular vertex
  is at most $\sum_{i=1}^{\infty} i p (1-p)^i = (1-p)/p$. As $e$
  contains $d$ vertices of $R$, the expected number of children of
  $v_e$ is at most $d (1-p)/p = 1/(2 - 1/d)$; since $d \ge 2$, the
  expected number of children is at most $2/3$. It follows that the
  expected size of a subtree rooted at $v_e$ is at most $3$.
  
  Note that $\Rev_2(e)$ and $\Rev_3(e)$ are both $0$ if $v_e$ is not
  the root of its tree in $F$. Conditioned on $v_e$ being a root,
  $\Exp{\Rev_2(e)} \le 3 w(e)$, as $e$ is the heaviest edge in its
  tree, and the expected size of the tree is at most
  $3$. $\Exp{\Rev_3(e)}$ is at least $w(e)/2$, as $e$ intersects no
  previously added edges, and with probability at least $1/2$, it
  intersects no edge added to $M_2$ later. Therefore, the ratio of
  these expectations is at most $6$, completing the proof.
\end{proof}

Now, we define our final algorithm {\sc SampleAndPrice} for the HVM
problem:

\begin{algo}
  \underline{\sc SampleAndPrice($|L|, R$)}\\
  $k \leftarrow Binom(|L|, 1-\frac{1}{2d})$ \\
  Let $L'$ be the first $k$ elements of $L$.\\
  $M_1 \leftarrow$ {\sc Greedy($G(L', R)$)}. \\
  For each $v \in V$: \+ \\
    Set $price(v)$ to be the weight of the edge incident to $v$ in
    $M_1$. \- \\
  $M \leftarrow \emptyset$ \\
  For each subsequent $l \in L - L'$: \+ \\
    Let $e$ be the highest-weight edge containing $l$ such that for
    each $v \in e$, $w(e) \ge price(v)$ \\
    If $e$ is disjoint from $M$, add $e$ to $M$. \-
\end{algo}

As before, since the input is a random permutation of $L$, $L'$ is a
subset of $L$ in which every vertex is selected independently with
probability $1 - 1/2d$, and the matching $M$ is at least as good as $M_3$
from {\sc Simulate}. Therefore, we have proved the following theorem:

\begin{theorem}\label{thm:HVM}
  {\sc SampleAndPrice} is an $O(d^2)$-competitive algorithm for the
  HVM secretary problem.
\end{theorem}

Note that $M$ may also contain extra edges that occur earlier in the
permutation than edges they intersect; for the BVM problem, this was
the difference between Lemma~\ref{lem:wM3} and the stronger bound
\ref{lem:tighterBound}. We do not provide a tighter analysis similar
to Lemma~\ref{lem:tighterBound} for the HVM problem in this extended
abstract, nor make an attempt to optimize the constants of
Lemma~\ref{lem:HVM}. In particular, for the HEM problem with $d=2$
(finding an online matching in a non-bipartite graph $G(V,E)$, given a
random permutatation of $E$), we have a constant bound on the
competitive ratio; a smaller constant can easily be obtained.

\begin{figure}
  \begin{center}
    \begin{tikzpicture}[yscale=0.75,xscale=1.25]
      \node (hvm) at (2,3) {HVM($d+1$)};
      \node (hem) at (0,1.5) {HEM($d$)};
      \node (bvm) at (4,1.5) {BVM};
      \node (bem) at (0,0) {Edge-at-a-time Matchings};
      \node at (0,-0.5) {(In arbitrary graphs)};
      \node (tvs) at (4,0) {Transversal Matroid Secretary};

      \draw[->] (hvm) -- (hem); \draw[->] (hvm) -- (bvm);
      \draw[->] (hem) -- (bem); \draw[->] (bvm) -- (tvs);
    \end{tikzpicture}
  \end{center}
  \vspace{-0.25in}
  \caption{Relationships between the HVM problem and various special cases.}
\end{figure}
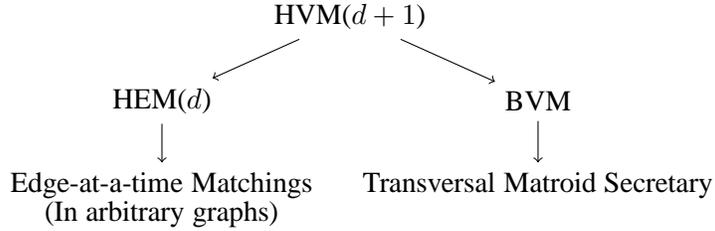

\section{Secretary Problems with Groups}
\label{sec:groups}

Consider a secretary-type problem in which, instead of receiving a
random permutation of the elements, elements can be grouped by an
adversary. The algorithm receives the number of groups in advance,
instead of the number of elements. However, once the groups have been
constructed, they arrive in random order; when a group arrives, the
algorithm can see all its elements at once.  Note that the groups are
fixed in advance; the adversary cannot construct groups in response to
the algorithm's choices or the set of groups seen so far. The effect
of such grouping on the difficulty of the problem is not immediately
clear: The adversary can ensure that some permutations of the element
set never occur, which might make the problem more difficult. On the
other hand, as the algorithm is allowed to see several elements at
once, it may be easier to compute a good solution.

For instance, consider the classical secretary problem with groups.
An optimal algorithm will never hire any but the best secretary from a
group, and it is easy to obtain an $e$-competitive algorithm: Ignore
all but the best secretary from each group, and run the standard
secretary algorithm on these. That is, observe a constant ($1/e$)
fraction of the groups, and note the value/price of the best secretary
seen so far. From the rest of the input, hire the best secretary from
the first group with a secretary to beat this price. Perhaps a reason
this problem is as easy as the original version is that only one
element is to be selected.

By way of contrast, consider the following matching problem, even
restricted to bipartite graphs: The algorithm is initially given the
vertex set of a bipartite graph, and an adversary groups the edges
arbitrarily. The groups arrive in random order; when a group arrives,
the algorithm sees the weights of all edges it contains. The goal is
to find a maximum-weight matching; note that as a special case of HEM
with $d=2$, we have an $O(1)$-competitive algorithm for
this problem without edge grouping. A natural Sample-And-Price
algorithm for this problem is as follows: Look at a constant fraction
of the input, and construct a matching with these edges (either the
optimal matching, or the greedy matchings we used in the previous
sections). Use the weights of edges in the matching to set vertex
prices, and in the remainder of the input, select an edge if its
weight is at least the price of each of its endpoints, and if it does
not conflict with edges already selected. Unfortunately, this
algorithm does not work: Consider a bipartite graph $G(L \cup R)$,
with $L = \{l_1, l_2, \ldots, l_n\}$ and $R = \{r_1, r_2, \ldots,
r_n\}$. We have two groups of edges: $E_1 = \{(l_i, r_i) | 1 \le i \le
n\}$, with $w((l_i, r_i)) = 1 + 2i\eps$, and $E_2 = \{(l_i, r_{i+1}) |
1 \le i < n \}$, with $w((l_i, r_{i+1})) = 1 + (2i+1)\eps$. Assuming
$\eps \ll 1/n^2$, $E_1$ corresponds to an optimal matching, with
weight $\approx n$. If $E_1$ arrives first, the price of each $r_i$ is
$1 + 2i\eps$. Subsequently, when $E_2$ arrives, $w((l_{i-1}, r_i))$ =
$1+ (2i-1)\eps$, and hence no edge of $E_2$ beats the price of its
right endpoint. If $E_2$ arrives first, the price of each $l_i$ is
$1+(2i+1)\eps$. Subsequently, when $E_1$ arrives, $w((l_i, r_i)) = 1 +
2i\eps$, and so no edge except $(l_n, r_n)$ beats the price of its
left endpoint, for a total revenue of $\approx 1$.

We believe, therefore, that the introduction of groups affects these
secretary-type problems in non-trivial ways, and these problems are
likely to be of theoretical interest; in addition, they have
applications to problems where groups occur naturally, and we do not
receive a random permutation of the entire element set. To take
another example from the advertising world, when a merchant plans a
campaign, she may submit to the reservation system multiple ads,
together with the slots in which each ad can be placed, and a price
for each ad-slot combination. Even if the merchants arrive in a random
order, this does not correspond to a random permutation of ads, and
hence our previous analysis is not directly applicable. We model this
(as in BVM) as an edge-weighted matching problem on a bipartite graph
$G(L \cup R, E)$ in which vertices of $L$ may be grouped; here, the
groups correspond to the set of ads for a given advertiser. The
algorithm initially receives $R$ (the set of slots), and the number of
advertisers/groups; the adversary can construct groups from $L$
arbitrarily. Once the groups have been fixed, a random permutation of
the groups is seen, and when a group arrives, the algorithm must
decide which ads to accept, and where to place them; as always,
decisions are irrevocable. We refer to this as the BVM problem with
groups. 

\begin{theorem}\label{thm:BVMlogn}
  There is an $O(\log n)$-competitive algorithm for the BVM problem
  with groups.
\end{theorem}
\vspace{-0.1in} 
It is easy to prove this theorem using standard techniques: Sample the
first half of the vertices, and let $w$ denote the weight of the
heaviest edge seen so far. Pick an integer $j$ uniformly at random in
$[0, 1 + \ceil{\log_2 n}]$, and set a threshold of $w/2^j$. In the
second half, greedily construct a matching using edges with weight
above the threshold. (See, for instance, Theorem 3.2 of
\cite{MatroidSec} for analysis of an essentially similar algorithm.)
For completeness, we give a proof of Theorem~\ref{thm:BVMlogn} in
Section~\ref{subsec:otherProofs} of the appendix.

\begin{figure}[h]
  \begin{center}
    \begin{tikzpicture}[xscale=0.75]
      \tikzstyle{vertex}=[circle,draw,inner sep=0pt,minimum size=6mm];
      
      \node (a) at (0,2) [vertex] {$A$}; \node (b) at (0,1) [vertex] {$B$};
      \node (c) at (0,0) [vertex] {$C$};

      \node (x) at (4,1.5) [vertex] {$X$}; \node (y) at (4,0.5) [vertex] {$Y$};

      \draw (a) -- (x); \node at (2,1.9) {4};
      \draw (b) -- (x); \node at (2,1.4) {3};
      \draw (b) -- (y); \node at (2,0.9) {2};
      \draw (c) -- (y); \node at (2,0.4) {1};

    \end{tikzpicture}
  \end{center}
  \caption{Example for BVM with groups. Vertices A,C are in group 1,
    and vertex B is in group 2. Using the {\sc SampleAndPrice}
    algorithm, if group 2 is sampled and group 1 is not, both edges
    incident to A and C beat their prices, and hence are added to
    $M_2$. If both groups are sampled, A will be matched to X and B to
    Y in $M_1$, while $C$ will remain unmatched.}\label{fig:2}
\end{figure}
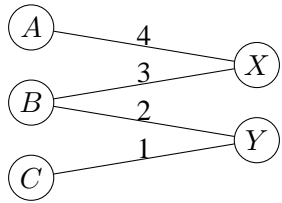

A natural question is whether one can find a constant-competitive
algorithm for BVM with groups. Note that one must be careful about
using Sample-And-Price algorithms: First, as the example above shows,
the natural algorithm with groups of edges instead of vertices does
not work. Second, one might sample a constant fraction of groups,
construct a matching $M_1$ on the sampled groups, and then use $M_1$
to set prices. However, once prices have been set in this way, the
edges assigned to a group $g$ may not be the same as the edges that
would have been assigned to $g$ in $M_1$ if $g$ had been sampled. This
was not the case for the basic BVM problem: If an edge $(l,r)$ is in
$M_2$, then by construction -- fixing all other coin flips -- if the
coin for $l$ had come up heads instead of tails, $(l,r)$ would be in
$M_1$. As the example in Figure~\ref{fig:2} shows, this desirable
property no longer holds once groups are introduced.

We conjecture that the following algorithm {\sc SampleWithGroups} is
constant-competitive for BVM with Groups. Here, $\script{G}$ denotes
the set of groups:

\begin{algo}
  \underline{\sc SampleWithGroups($|\script{G}|, R$)}\\
  Sample each group with probability $p$.\\
  Construct a greedy matching $M_1$ on the set of sampled groups
  $\script{G'}$.\\
  $M_2 \leftarrow \emptyset$. \\
  For each group $g$ in $\script{G} - \script{G'}$: \+ \\
    Let $E'$ denote the edges assigned to vertices of $g$ in the
    greedy matching on $\script{G'} \cup g$. \\
    $M_2 \leftarrow M_2 \cup E'$. \- \\
  $M_3 \leftarrow M_2$ \\
  For each $r \in R$: \+ \\
    If $r$ has degree $> 1$ in $M_3$: \+ \\
      Delete all edges incident to $r$ from $M_3$.
\end{algo}

It is easy to see that $\Exp{w(M_1)} \ge p \opt/2$. By construction,
the edges assigned to $g$ in $M_2$ are precisely those that would have
been assigned to $g$ in $M_1$ if $g$ had been sampled. (Hence, this
algorithm differs from the natural {\sc SampleAndPrice}.) Therefore,
it follows that the probability an edge contributes to $M_2$ is
$(1-p)/p$ times the probability it contributes to $M_1$. If $p = 1/2$,
it follows that $\Exp{w(M_2)} = \Exp{w(M_1)}$, and further, that the
expected degree of $r \in R$ in $M_2$ is equal to its expected degree
in $M_1$, which is at most 1 since $M_1$ is a matching. This does not
suffice to give a lower bound on the expected weight of $M_3$, but we
conjecture that the expected weight of $M_3$ is at most a constant
factor lower than that of $M_2$.

\begin{conjecture}
  {\sc SampleWithGroups} is constant-competitive for the BVM problem
  with groups.
\end{conjecture}

\section{Graphic Matroids}
\label{sec:graphic}

In this section, we describe a $2e$-competitive algorithm for the
Graphic Matroid Secretary problem. Here, we are initially given the
set of vertices $V$ of an undirected edge-weighted graph $G=(V,E)$
together with the size of its edge set $|E|$.  The edges of the graph
appear in a random order, and the goal is to accept a maximum-weight
subset of edges $F$ that does not contain any cycles.  As always, the
decision to accept an edge must be made upon its arrival, and cannot
be revoked.

This problem is equivalent to finding the maximum-weight spanning tree
(assuming $G$ is connected) and is also equivalent to finding the
maximum-weight independent set in the graphic matroid defined by the
graph $G$.  Babaioff et al. \cite{MatroidSec} give a 16-competitive
algorithm for the secretary version of this problem based on a related
algorithm for transversal matroids.
We give a simple reduction to the classical secretary problem, losing
a factor of $2$ in the reduction. In this way, we obtain a $2e
\approx5.436$-competitive algorithm for the Graphic Matroid Secretary
problem.

Fix an ordering $v_1,v_2,\dots,v_n$ on the vertices of $G$.  Consider
two directed graphs: graph $G_0$ is obtained by orienting every edge
of $G$ from higher numbered to lower numbered vertex, and graph $G_1$
by orienting every edge from lower to higher numbered vertex.

Our online algorithm initially flips a fair coin $X\in \{0,1\}$.  For
each vertex $v$ independently, it runs a secretary algorithm to find
the maximum-weight edge leaving $v$ in $G_X$. The output of the
algorithm is $F'$, the union of all edges accepted by the individual
secretary algorithms. Since the graph $G_X$ is acyclic and each vertex
has at most one outgoing edge, the set of edges $F'$ must be acyclic
even in the undirected sense.

It remains to show a lower bound on the weight of $F'$.  For each
vertex $v$, let $h_X(v)$ be the heaviest edge leaving vertex $v$ in
$G_X$. Let $F_X = \{h_X(v) ~|~ v\in V\}$.  Let $F^*$ be a
maximum-weight acyclic subgraph of $G$.

\begin{prop} \label{prop:dir-tree}
  $\sum_{v\in V} w(h_0(v)) + w(h_1(v)) \ge \sum_{e\in F^*} w(e)$.
\end{prop}

Conditioned on the coin flip $X$, each secretary algorithm recovers at
least $1/e$ fraction of the weight of the heaviest edge leaving its
vertex. Hence $E[w(F') ~|~ X=x] = \frac{1}{e} w(F_x)$ for $x=0,1$.
Using Proposition \ref{prop:dir-tree}, $E[w(F')] = \frac{1}{e} \left(
  \frac12 E[w(F') ~|~ X=0] + \frac12 E[w(F') ~|~ X=1] \right) \ge
\frac{1}{2e}w(F^*)$. Therefore, we obtain the following theorem:

\begin{theorem}
  There is a $2e$-competitive algorithm for the graphic matroid
  secretary problem.
\end{theorem}

\section{Conclusions and Open Problems}

We list several problems that remain to be solved:
\begin{itemize}

\item An improved understanding of groups -- and their contribution to
  the difficulty of secretary-type problems -- is likely to be of
  interest. In particular, it may be possible to find a
  constant-competitive algorithm for the BVM problem with groups.

\item Few lower bounds for these problems are known beyond $1/e$ for
  the original secretary problem; obtaining such bounds may require
  new techniques.

\item In the basic BVM problem, we lose a factor of 2 by constructing
  greedy matchings. If, instead, we modified our algorithm to set
  prices using an optimal matching $M_1$ on the sampled vertices, is
  the resulting algorithm 4-competitive? Is it even $O(1)$-competitive?

\item Finally, obtaining an $O(1)$-competitive algorithm for the
  general matroid secretary problem is still open, though the
  competitive ratios for important special cases such as transversal
  and graphic matroids have been reduced to small constants.
\end{itemize}

\medskip
\noindent
\textbf{Acknowledgments:} We would like to thank Florin Constantin,
Jon Feldman, and S. Muthukrishnan for helpful discussions on BVM and
related problems.

\bibliographystyle{Plain}

\appendix
\section{Omitted Proofs}

\subsection{Proof of Lemma~\ref{lem:tighterBound}}

We prove Lemma~\ref{lem:tighterBound} below, showing that {\sc
  SampleAndPrice} is 8-competitive for the BVM problem.

For each $v \in R$, we let $\Rev_2(v)$ be the revenue earned by $v$ in
$M_2$, which is the total weight of edges in $M_2$ incident to
$v$. Similarly, $\Rev_3(v)$ denotes the weight of the (at most one)
edge of $M_3$ incident to $v$. Let $P_i$ be the probability that $v$
is incident to $i$ edges in $M_2$. Finally, we let $\Exp{\Rev_2(v)|i}$
and $\Exp{\Rev_3(v)|i}$ be the expected revenue earned by $v$ in $M_2$
and $M_3$ respectively, conditioned on $v$ being incident to $i$ edges
in $M_2$.

First, we note that $\Exp{\Rev_3(v)|i} = \frac{\Exp{\Rev_2(v)|i}}{i}$,
as for each set of coin flips in which $v$ has degree $i$ in $M_2$, we
may see any of the $i$ edges incident to $v$ first in the random
permutation; on average, then, we receive a $1/i$ fraction of
$\Rev_2(v)$. We then have the following equations:

\begin{eqnarray}
  \Exp{\Rev_2(v)} &=& \sum_{i=1}^{\infty} P_i \cdot \Exp{\Rev_2(v)|i}.\\
  \Exp{\Rev_3(v)} &=& \sum_{i=1}^{\infty} P_i \cdot \frac{\Exp{\Rev_2(v)|i}}{i}.
\end{eqnarray}

For ease of notation below, we use $w_i$ to denote
$\Exp{\Rev_2(v)|i}$. We wish to bound $\Exp{\Rev_3(v)}$ in terms of
$\Exp{\Rev_2(v)}$, and we do this as follows: First, we show that $P_i
\le (1-p) P_{i-1}$, and $w_i \le \frac{i}{i-1} w_{i-1}$. Next, we
prove that subject to these constraints, the worst-case ratio of these
two expectations occurs when all the constraints hold with
equality. We can then evaluate the sums, and show that
$\Exp{\Rev_3(v)} \ge p \Exp{\Rev_2(v)}$, completing our proof.

It is easy to see that $w_i \le \frac{i}{i-1} w_{i-1}$; consider any
partial history of {\sc Simulate} in which $i-1$ edges incident to $v$
have been added to $M_2$ so far; as we process edges in decreasing
order of weight, the $i$th edge must be the lightest of those seen so
far. As this is true for each (partial) history, it holds in
expectation, and so $w_i \le w_{i-1} + \frac{w_{i-1}}{i-1}$.
Similarly, to see that $P_{i} \le (1-p) P_{i-1}$, consider a partial
history until the $(i-1)$st edge has just been added: $M_2$ will have
$i-1$ edges incident to $v$ if the coin for the next edge incident to
$v$ considered by {\sc Simulate} comes up ``heads'', with probability
$p$. $M_2$ will have $i$ edges incident to $v$ if the coin for the
next edge incident to $v$ comes up ``tails'', and that for the
following edge comes up heads, with probability $(1-p)\cdot
p$.\footnote{It is possible that there is only one more edge incident
  to $v$, in which case $v$ will have $i$ edges with probability
  $(1-p)$. However, this only helps the analysis. Alternatively, one
  can assume the existence of a large number of ``zero-weight'' edges
  incident to $v$.}  Again, as this holds for each history, we have
$P_i \le (1-p) P_{i-1}$.

To see that the worst-case ratio occurs when all these constraints
hold with equality, notice that the ratio between successive terms of
Equations (1) and (2) is increasing: The ratio between the $i$th terms
is simply $i$. Let $\alpha$ denote the worst-case ratio of the
expectations; from Lemma~\ref{lem:wM3}, we already know that $\alpha
\le p^2$. If $j = \floor{1/\alpha}$, for $1 \le i \le j$, the ratio
between the $i$th term of the two sums is at most $\alpha$, while for
$i > j$, the ratio is greater than $\alpha$. Consider a choice of
$w_i$'s and $P_i$'s such that the ratio between (1) and (2) be as
large as possible, and suppose the constraints on $P_i$ and $w_i$ do
not all hold with equality. Let $k$ be an index such that $P_k < (1-p)
P_{k-1}$ or $w_k < \frac{k}{k-1} w_{k-1}$. If $k > j$, then by
increasing $P_k$ or $w_k$, we do not violate any constraint, and the
increase in (1) is greater than $\alpha$ times the increase in
(2). Similarly, if $k \le j$, by decreasing $P_{k-1}$ or $w_{k-1}$ to
achieve equality, and also decreasing $P_1 \ldots P_{k-2}$ or $w_1
\ldots w_{k-2}$ to maintain feasibility, the decrease in (1) is less
than $\alpha$ times the decrease in (2). In either of these
situations, we increase the ratio between the two sums, contradicting
our initial setting of $w_i, P_i$.

Finally, we can now evaluate this worst case ratio. Setting $w_{i} =
\frac{i}{i-1} w_i$ and $P_i = (1-p) P_{i-1}$, we find:
\begin{eqnarray*}
  \Exp{\Rev_2(v)} &=& \sum_{i=1}^{\infty} i w_1 P_1 (1-p)^{i-1} = w_1 P_1
  / p^2. \\
  \Exp{\Rev_3(v)} &=& \sum_{i=1}^{\infty} w_1 P_1 (1-p)^{i-1} = w_1
  P_1 / p = p \Exp{\Rev_2(v)}\\
\end{eqnarray*}
As $\sum_v \Exp{\Rev_2(v)} = \Exp{w(M_2)} \ge (1-p) \opt/2$, we have
$\Exp{w(M_3)} \ge p (1-p) \opt/2$, completing
the proof of Lemma~\ref{lem:tighterBound}.





\subsection{Other Proofs}\label{subsec:otherProofs}

\begin{proofof}{Theorem~\ref{thm:BVMlogn}}
  We show that the algorithm of Theorem~\ref{thm:BVMlogn} is $O(\log
  n)$-competitive for BVM with groups, closely following the analysis
  of \cite{MatroidSec} for an $O(\log k)$-competitive algorithm for
  general matroids. Recall that the algorithm observes the first half
  of the vertices, and picks a random integer $j \in [0, 1 +
  \ceil{\log n}]$. If $w$ is the weight of the heaviest edge seen so
  far, the algorithm sets a threshold of $w/2^j$, and in the second
  half, greedily constructs a matching using edges of weight greater
  than this threshold.

  Let $\opt$ be an optimal matching; we also abuse notation and use
  $\opt$ to refer to the weight of this matching, though the meaning
  will be clear from context.  Let $w_1, w_2, \ldots w_k$ denote the
  weights of edges in $\opt$, such that $w_i \ge w_{i+1}$
  for $1 \le i < k$. Let $q$ denote the largest index in $[1,k]$ such
  that $w_q \ge w_1/n$.  Clearly, $\sum_{i=1}^q w_i > \opt/2$, as
  the remaining edges all have weight less than $w_1/n$, and there are
  fewer than $n$ of them. For any set of edges $F$, we use $n_i(F)$ to
  denote the number of edges in $F$ with weight at least $w_i$, and
  $m_i(F)$ to denote the number of edges in $F$ with weight at least
  $w_i/2$. Now, we have:
  \[ \sum_{i=1}^q w_i = \left(\sum_{i=1}^{q-1} n_i(OPT) (w_i - w_{i+1})\right) +
  n_q(OPT) w_q \]

  Let $M$ be the matching returned by our algorithm. We lower bound
  the weight of $M$ as follows:
  \[ w(M) \ge \frac{1}{2} \left(\sum_{i=1}^{q-1} m_i(M) (w_i -
    w_{i+1}) \right) + \frac{m_q(M) w_q}{2} \]
  
  In order to obtain an $O(\log n)$-competitive algorithm, it suffices
  to show that for each $1 \le i \le q$, $\Exp{m_i(M)} \ge
  n_i(\opt)/O(\log n)$. First, consider the case of $i=1$: $n_1(\opt)
  = 1$, and we argue that $\Exp{m_1(M)} \ge 1/4 (\ceil{\log n} +
  1)$. With probability $1/4$, the vertex $v$ incident to the heaviest
  edge appears in the second half, and the heaviest edge not incident
  to any vertex of $v$'s group appears in the first half. If this
  occurs, and the algorithm picks $j=0$ (which happens with
  probability $1/(\ceil{\log n} + 1)$), then the only edges with
  weight above the threshold are those incident to vertices in $v$'s
  group. Thereore, the greedy algorithm will select the heaviest edge
  with probability $1/4 (\ceil{\log n} + 1)$, and hence $\Exp{m_1(M)}
  \ge 1/4 (\ceil{\log n} + 1)$.

  We now complete the argument for each $i > 1$. Let $v$ be the vertex
  incident to the heaviest edge. We consider two cases: First, that
  at least half the edges of $\opt$ with weight at least $w_i$ are
  incident to vertices not in the same group as $v$, and second, that
  more than half these edges are incident to vertices of $v$'s group.

  In the former case, suppose that $v$ is seen in the first half. Let
  $w$ be the weight of this heaviest edge, and let $i'$ be the
  smallest integer in $[0, 1+\ceil{\log n}]$ such that $w/2^{i'} \le
  w_i$.\footnote{Note that $w$ may be greater than $w_1$, as the
    heaviest edge may not be in $opt$. However, it is easy to see that
    $w \le 2w_1$, and since $w_i \ge w_1/n$, there always exists such
    an index $i'$.} With probability $\frac{1}{(\ceil{\log n} + 1)}$,
  the algorithm picks $j=i'$, and the threshold is set to be $w/2^{i'}
  > w_i/2$. Let $X$ denote the event that the threshold is set to be
  $w/2^{i'}$; as we have seen, $\Pr[X] \ge 1/2 (\ceil{\log n} +
  1)$. We show that conditioned on $X$, $\Exp{m_i(M)}$ is sufficiently
  large.
  
  Recall that $\opt$ contains a matching of size $i$ using edges of
  weight at least $w_i$; it follows that in expectation, using edges
  of this weight, there is a matching in the second half of size at
  least $i/4$. (This is because at least half of these $i$ edges are
  in other groups; even conditioned on $v$ appearing in the first
  half, each of the remaining $\ge i/2$ edges could appear in either
  half.)  Since we construct a greedy matching using edges of weight
  at least $w_i/2$, the expected size of this matching is at least
  $i/8$. Hence, with probability at least $\frac{1}{2 (\ceil{\log n} +
    1)}$, $\Exp{m_i(M)} \ge i/8 $. That is, $\Exp{m_i(M)} \ge i/16
  (\ceil{\log n} + 1)$.

  We now consider the second case, when more than half the edges of
  $\opt$ with weight at least $w_i$ are in the same group as $v$. Let
  $u$ be the vertex outside this group incident to the heaviest-weight
  edge. Suppose $u$'s group appears in the first half, and $v$'s group
  in the second. Let $w$ be the weight of the heaviest edge incident
  to $u$; if $w \le w_i$ and we pick $j = 0$, the \emph{only} edges
  above the threshold will be vertices in $v$'s group. Since we
  construct the greedy matching using only the group of $v$, and there
  exists a matching in this group with more than $i/2$ edges of weight
  $w_i$, the matching we construct has at least $i/4$ edges of weight
  at least $w_i$. If $w > w_i$, then with probability $1/\ceil{\log n}
  + 1$, we pick an index $j$ such that $w_i \ge w/2^{j} >
  w_i/2$. Again, we will find a matching in which at least $i/4$ edges
  have weight at least $w_i/2$. Therefore, with probability at least
  $\frac{1}{4 (\ceil{\log n + 1})}$, we find a matching of size at
  least $i/4$. Therefore, $\Exp{m_i(M)} \ge i/16 (\ceil{\log n} + 1)$.

  Therefore, we have $\Exp{w(M)} \ge \frac{1}{2} \frac{1}{16
    (\ceil{\log n} + 1)} \sum_{i=1}^q w_i \ge \opt/64 (\ceil{\log n} +
  1)$.
\end{proofof}

\bigskip
\begin{proofof}{Proposition~\ref{prop:dir-tree}}
  Let $h(v)$ denote the heaviest edge incident to $v$; clearly
  $\sum_{v} w(h_0(v)) + w(h_1(v)) \ge \sum_v w(h(v))$. It remains to
  show that this latter sum is at least $\sum_{e \in F^*} w(e)$. To
  see this, consider the tree $F^*$, and root it arbitrarily. For each
  edge $e = (u,v) \in F^*$, the weight of $e$ is at most $h(v)$, where
  $v$ is the vertex further from the root. Each vertex $v$ is charged
  by at most one edge, and so $\sum_v w(h(v)) \ge \sum_{e \in
    F^*} w(e)$. 
\end{proofof}
\end{document}